\documentstyle[prd,aps,maps,twocolumn,epsf]{revtex}

\begin{document}
\title{Can a ``natural'' three-generation neutrino 
mixing scheme satisfy everything?}
\draft
\author{Christian Y. Cardall and George M. Fuller}
\address{
Department of Physics, University of California, San Diego,
La Jolla, California 92093-0319}
\date{\today}

\mabstract{
We examine the potential for a ``natural'' three-neutrino
mixing scheme to satisfy available data and astrophysical 
arguments. 
By ``natural'' we mean no sterile neutrinos, and
a neutrino mass hierarchy similar to that of the charged leptons. 
We seek to
satisfy (or solve):
  1. Accelerator and reactor neutrino oscillation constraints, 
including LSND;
  2. The atmospheric muon neutrino deficit problem;
  3. The solar neutrino problem;
  4. Supernova $r$-process nucleosynthesis in neutrino-heated 
supernova ejecta;
  5. Cold+hot dark matter models.
We argue that putative supernova $r$-process nucleosynthesis 
bounds on
two-neutrino flavor mixing can be applied directly to 
three-neutrino mixing
in the case where one vacuum neutrino mass eigenvalue 
difference dominates the others.
We show that in this ``one mass scale dominance'' limit, a 
natural three-neutrino oscillation solution meeting all the above 
constraints exists only if the atmospheric neutrino data {\em and}
the LSND data can be explained with one neutrino mass 
difference.
In this model, an explanation for the solar
neutrino data can be effected by employing  the {\em other} 
independent 
neutrino mass difference.
Such a solution is only marginally allowed by the current data, 
and proposed long-baseline neutrino oscillation
experiments can definitively rule  it out. If it were ruled out,
the simultaneous solution of the above
constraints by neutrino oscillations would then require 
sterile neutrinos
and/or a neutrino mass hierarchy of a different nature than
 that of the charged leptons. \\ \\
PACS number(s): 14.60.Pq, 96.40.Tv, 97.60.Bw
}

\mmaketitle
\narrowtext

% body of paper here

\section{Introduction}
\label{sec:intro}

Neutrinos with masses and mixings are one of the 
simplest extensions of
the standard model of elementary particles. All direct searches for 
neutrino oscillations using neutrino beams from accelerators 
and reactors
have produced only upper limits, with the possible exception 
of the Liquid
Scintillator Neutrino Detector experiment (LSND) at 
Los Alamos \cite{lsnd}. The claimed LSND signal is still
somewhat controversial; in Ref. \cite{hill95} the LSND data
are interpreted as yielding only an upper limit. However, 
for the purposes
of this paper we will accept the interpretation of the
LSND data as a neutrino oscillation signal, except where
explicitly stated otherwise.

In spite of a dearth of direct evidence, neutrino mixing is a 
popular explanation for a number of
measured neutrino phenomena that appear to be at 
variance with predictions
based on massless, non-mixing neutrinos. These phenomena  
include the so-called solar
and atmospheric neutrino ``deficits'' (\cite{solrev} and \cite{atrev}, 
respectively). Massive neutrinos can affect cosmological evolution
\cite{omeganu} and large-scale structure formation 
\cite{lssrev,primack95}.
Neutrino flavor mixing could affect
supernova dynamics \cite{fuller87,fuller92} 
and nucleosynthesis \cite{qian93,qian95,sigl95}, 
and big bang nucleosynthesis \cite{enqvist92,shi93}. 
These cosmological/astrophysical settings
sometimes suggest stricter limits on neutrino masses 
and mixings than those 
obtainable with earth-based experiments.

For simplicity, neutrino oscillations are often analyzed 
in a two-flavor
framework, with one neutrino mass eigenvalue difference 
and one mixing angle.
In the wake of the LSND result, 
interest has grown in constructing models of neutrino mixing that
accommodate ``everything.'' In several of these models, 
results from
two-flavor interpretations of various physical effects are combined
to make a consistent composite model 
\cite{neumodels,raffelt95,chun95}. In addition, 
some of these models use
three independent mass differences: one each for solar neutrinos,
atmospheric neutrinos, and LSND.

Two aspects of these models might be considered
``unnatural.'' First, the use of three independent  
neutrino mass eigenvalue differences
requires the introduction of a fourth neutrino. In light 
of measurements of the width
of the $Z_0$ at LEP \cite{zwidth}, this fourth neutrino 
must be taken to be 
``sterile'' (an $SU(2)$ singlet).
Second, these models sometimes employ 
an ``inverted'' neutrino mass hierarchy. In these 
inverted schemes, the neutrino  
mass eigenvalue most closely associated with $\nu_e$ is heavier
than those associated with $\nu_{\mu}$ or $\nu_{\tau}$, or the  
mass eigenvalue most closely associated with 
$\nu_{\mu}$ is heavier
than that associated with $\nu_{\tau}$. 

One might hope that these unnatural features could be 
removed by employing
a genuine three-neutrino mixing scheme.
It is apparent that non-trivial mixing among three or 
more generations 
of neutrinos has the possibility of a richer phenomenology 
than models in
which two-generation mixings are ``stitched together.'' 
In particular, the 
excess of electron- over muon-induced events observed 
in atmospheric neutrinos could
be due to the $\nu_{\mu}$ oscillating into {\it both} 
$\nu_e$ and $\nu_{\tau}$,
not just one or the other. 

As noted earlier, several previous models have used 
three independent 
neutrino mass eigenvalue differences. However, 
a three generation scheme 
has only two independent mass differences, which we 
label $\delta m^2_1$ 
and $\delta m^2_2$. We define $\delta m^2$ to be difference
of the squares of two vacuum neutrino mass eigenvalues, and 
take $\delta m^2 > 0$. 
In the Mikheyev-Smirnov-Wolfenstein
(MSW) mechanism \cite{msw,bethe},  matter effects 
can enhance or suppress 
neutrino mixing and may lead to flavor conversion.
This mechanism is a popular explanation of the solar 
neutrino problem. 
Since a mass-level crossing is
the basis of the MSW effect, we are 
forced to take one of the independent neutrino mass 
differences from 
a relatively narrow range determined by solar parameters. 
An MSW solution 
to the solar neutrino problem thus determines, 
for example, $\delta m^2_1$.

Note, however, that a fair range of neutrino mass 
differences can be employed
in vacuum oscillation explanations of 
the LSND and atmospheric
neutrino data.
In a ``last resort'' effort to find a natural three-neutrino 
mixing scheme,
we here consider the possibility of explaining both the
atmospheric neutrino data and the LSND data 
with the other independent neutrino mass difference, 
$\delta m_2^2$.
Unfortunately, this common neutrino mass eigenvalue 
difference would have to
lie in the range $\delta m_2^2 \approx 0.2-0.4$~eV$^2$. 
This particular range of values for $\delta m_2^2$ would 
be an order of 
magnitude {\em lower} than the neutrino mass difference 
 most commonly associated with LSND, while it would be 
an order of
magnitude {\em larger} than the most popular value 
associated with atmospheric 
neutrinos. Further, this single common neutrino mass 
eigenvalue difference 
turns out to be narrowly specified: as we shall see, it is
bounded from above by supernova $r$-process considerations 
(for non-inverted neutrino mass hierarchies), and 
more strictly by
the compatibility of atmospheric neutrinos with 
laboratory limits; and
from below by the compatibility of LSND with other 
laboratory limits.

Is the use of the same neutrino mass difference ($\delta m_2^2$) 
 to give a vacuum neutrino oscillation
solution to {\em both} the 
LSND data and the published atmospheric neutrino results 
warranted? 
A wide range of neutrino mass differences appears to be 
capable of providing a neutrino oscillation explanation of the 
LSND data (see Fig. 3 of Ref. \cite{lsnd}).
With one of the neutrino mass eigenvalues set to zero, 
the ``favored'' 
LSND value
of $\delta m^2 \approx 6$ eV$^2$ \cite{primack95} yields 
neutrino masses that
are convenient from the perspective of cold+hot dark 
matter models.  
Note, however,
that neutrino oscillation interpretations of the 
LSND data only probe mass {\em differences}. We can 
compensate for a smaller 
mass difference, $\delta m_2^2 \approx 0.2-0.4$ eV$^2$, 
by offsetting 
{\em all} of the neutrino mass eigenvalues from zero
\cite{neumodels,raffelt95}. Such an offset would allow 
the {\em sum} of the
neutrino mass eigenvalues to provide the requisite 
contribution of hot dark matter
in the models of Ref. \cite{primack95}. Furthermore,
the ($\sin^2 2\theta$, $\delta m^2$) plot of the allowed LSND 
oscillation parameters \cite{lsnd} shows that 
compatibility with KARMEN 
\cite{armbruster95}, BNL E776 \cite{borodovsky92}, and
CCFR \cite{mcfarland95} 
is readily achieved for
$\delta m_2^2 \approx 0.2-0.4$~eV$^2$ \cite{lsnd}.

The range of $\delta m_2^2$ allowed by a neutrino 
oscillation solution to the
 atmospheric neutrino anomaly is a more 
subtle issue. The isotropy of the sub-GeV data 
makes them amenable to 
an oscillation solution for any 
$\delta m_2^2 \gtrsim 10^{-3}$ eV$^2$ (see e.g. \cite{acker94} and
references therein). At one point it was argued that 
data from upward-going 
muons restricted $\delta m_2^2$ to values around 
$10^{-2}$ eV$^2$ \cite{becker92}. However, this 
conclusion relied on calculations of
{\em absolute} neutrino fluxes and 
cross-sections---not just ratios of
these quantities, as can be used in analyzing contained events. It
has now been pointed out that calculations of 
absolute neutrino fluxes and
cross-sections different from those used in Ref. \cite{becker92}  
permit all of the parameter space allowed by the 
sub-GeV contained events,
in particular ``high'' ($\sim 10^{-1}$eV$^2$) 
values of $\delta m_2^2$ \cite{frati93,atrev}. 

Potentially more damaging to ``high'' $\delta m_2^2$ values 
in this context are the 
Kamiokande multi-GeV data, especially the claim 
that the data show
zenith angle dependance \cite{fukuda94}. The 
Kamiokande group's best
fit to this data would imply $\delta m_2^2 
\approx 10^{-2}$ eV$^2$, with 90\% C.L.
upper limits at $\delta m_2^2 \sim 0.1$ eV$^2$.  We note that 95\% 
C.L. contours could extend the allowed range of mass differences
to $\delta m_2^2 \sim 0.3$ eV$^2$. In addition, the statistical
significance of the Kamiokande group's best fit has 
been questioned 
\cite{saltzberg95,fogli295}. 

We will now explore what would be possible
if new analyses of the atmospheric neutrino data, or 
future data with better
statistics, were to allow $\delta m_2^2 \sim 0.3$ eV$^2$.
By assigning the other independent mass difference to be 
$\delta m_1^2 \sim 10^{-5}$ eV$^2$ to use for solar
neutrinos, our scheme falls into the category that the authors of
Ref. \cite{fogli95} call ``one mass
scale dominance'' (OMSD), in which $\Delta_{32}, \,
 \Delta_{31} \gg \Delta_{21}$ (where $\Delta_{ij} 
\equiv |m^2_i - m^2_j|$,
and $m_i$ and $m_j$ are neutrino mass eigenvalues).
Great simplification occurs in this limit \cite{omsd}.
Here we take the vacuum mass eigenstates 1, 2, and 3 to 
be those most closely corresponding to
the electron, muon, and tau neutrino flavor eigenstates, 
respectively.

In Ref. \cite{fogli95} the OMSD limit is used for 
neutrinos propagating in vacuum, and
it is shown that interpretations of experiments based 
on this scheme can 
easily be related to two-flavor interpretations.  
Additionally, Ref. \cite{fogli95}
offers new interpretations of 
the available accelerator and reactor data in terms of this simple 
three-generation framework. In this OMSD scheme, the 
CP-violating phase, which is inherent in a 
three-neutrino mixing framework,
and the mixing angle $\theta_{12}$ drop out of the 
problem. With this simplification,  neutrino
oscillation effects in vacuum can be described in 
terms of the two mixing angles 
$\theta_{13},\,\theta_{23}$ and
one mass-squared difference, $\Delta \equiv 
\Delta_{32} \approx \Delta_{31}$.

Previous authors have studied the three-flavor MSW effect in
the limit of well-separated mass scales \cite{kuo87}. For the 
matter density scales relevant
to the solar neutrino problem, a decoupling to an 
effective two-flavor mixing 
problem occurs, allowing an MSW solution employing only 
$\theta_{12}$ and $\Delta_{12}$.
For density scales relevant to supernovae, a 
similar decoupling occurs, leading
to an effective two-flavor mixing described in 
terms of $\theta_{13}$ and
$\Delta$. In both of these cases, the CP-violating 
phase does not appear in the
final results.

In Sec.\ \ref{sec:super} we argue that,
for the supernova hot-bubble/$r$-process environment, 
the survival probability
$P(\nu_e \rightarrow \nu_e)$ is roughly the same in the 
OMSD three-neutrino mixing case as in the 
two-neutrino mixing case. This allows the general 
results of calculations of the 
effects of two-neutrino mixing on the
supernova $r$-process to be directly applied to OMSD 
three-neutrino mixing. In Sec.\ 
\ref{sec:solution}
we discuss the chances for obtaining a natural solution 
to ``everything.'' 
Concluding remarks, along
with a discussion of the prospects of future experiments
to clarify the issues discussed in this paper, are contained in 
Sec.\ \ref{sec:concl}.

\section{Matter-enhanced three-neutrino mixing in the
 OMSD limit in supernovae}
\label{sec:super}

We now examine matter effects on the propagation of 
neutrinos in the
OMSD three-neutrino mixing case in the post core-bounce 
supernova environment. 
First, we will consider only adiabatic resonant conversion. 
In this analysis,
we will 
ignore neutrino-neutrino forward scattering effects as well. 
We will comment on nonadiabatic neutrino state evolution 
and the effects of
neutrino-neutrino forward scattering at the end of this section.

An important quantity for the $r$-process in neutrino-heated
supernova ejecta is $P(\nu_e \rightarrow \nu_e)$,
the probability that a $\nu_e$ emitted from the neutrinosphere
will still be a $\nu_e$ at the ``weak freeze-out radius.'' 
The weak freeze-out 
radius is the distance
from the center of the nascent neutron star at which the 
weak reactions
freeze out of equilibrium. 
Above the weak freeze-out radius the neutron-to-proton ratio,
$n/p$, can be taken as essentially fixed.

The $\nu_e$ survival probability is crucial, since
the average energy of the $\nu_e$ population can be altered by
a resonant flavor conversion involving either the 
$\nu_{\mu}$ or $\nu_{\tau}$ 
populations. This follows on noting that the average 
energies of the $\nu_{\mu}$
and $\nu_{\tau}$, $\langle E_{\nu_{\mu}, \nu_{\tau}} \rangle$, 
are always larger
than the average energy of the $\nu_e$, $\langle E_{\nu_e} 
\rangle$, in the 
absence of flavor conversion.
Following Refs. \cite{qian93} and \cite{loreti96}, 
we can approximate the effects
of neutrino flavor conversion on $\langle E_{\nu_e} \rangle$ as, 
\begin{eqnarray}
{\langle E_{\nu_e} \rangle}_{\rm{WFO}} = 
  &&P(\nu_e \rightarrow \nu_e)
   {\langle E_{\nu_e} \rangle}_{\rm{NS}}\nonumber \\
   && + \left[ 1 - P(\nu_e \rightarrow \nu_e)
   \right] {\langle E_{\nu_{\mu}, \nu_{\tau}} \rangle}_{\rm{NS}}. 
\label{ebar}
\end{eqnarray}
Here ``WFO'' stands for the weak freeze-out radius, 
and ``NS'' stands
for the radius of the ``neutrinosphere.'' A more 
complete discussion
of these issues can be found in 
Refs. \cite{qian93}
and \cite{loreti96}. 

As long as $P(\nu_e \rightarrow \nu_e)$ is
the same for both the two- and three-neutrino mixing 
cases, the impact 
of significant resonant flavor conversion on the average 
energy of the electron neutrino population (and hence 
the $n/p$ ratio)
will be the same, since the $\mu$ and $\tau$ 
neutrinos have a common energy. To be able to apply 
the results of 
calculations of the effects of two-neutrino mixing to the OMSD 
three-neutrino case, we need to show that, 
\begin{equation}
 P(\nu_e \rightarrow \nu_e)_{3-\rm{flavor,\,OMSD}} \approx
 P(\nu_e \rightarrow \nu_e)_{2-\rm{flavor}}.
\end{equation}

The neutrino amplitude propagation equation in the mass basis is,
\begin{equation}
i {d \over dx} \pmatrix{\nu_1 \cr \nu_2 \cr \nu_3}
   = {1 \over 2E} {\hat M}^2
   \pmatrix{\nu_1 \cr \nu_2 \cr \nu_3}, \label{neuprop1}
\end{equation}
where $E$ is the energy of the neutrino, and $x$ is a 
time development
parameter (e.g., radius).
The evolution matrix is
${\hat M}^2 / 2E$, where we take,
\begin{equation}
{\hat M}^2 = \pmatrix{ m_1^2 & 0 & 0 \cr 0 & m_2^2 & 0 
\cr 0 & 0& m_3^2},
\end{equation}
with $m_1$, $m_2$, and $m_3$ the neutrino mass eigenvalues.
In Eq. (\ref{neuprop1}), $\nu_i$ is the amplitude 
for the neutrino to be
found in mass eigenstate $i$, with $i=$1, 2, or 3.
Since the part of ${\hat M}^2$ proportional to the identity 
matrix contributes only
a universal phase, we may remove it and rewrite Eq. 
(\ref{neuprop1}) as,
\begin{equation}
i {d \over dx}\! \pmatrix{\!\nu_1\! \cr \!\nu_2\! \cr \!\nu_3\!}\!
   = \!{1 \over 6E}\! \pmatrix{\!\!-\Delta_{21}\!\! - \!\!\Delta_{31}\!\! 
   &0 &0 \cr 0 &
     \!\!\Delta_{21}\!\! - \!\!\Delta_{32}\!\! & 0 \cr 0&0& 
   \!\!\Delta_{32}\!\! +\!\! 
   \Delta_{31}\!\! }\!
   \pmatrix{\!\nu_1\! \cr \!\nu_2\! \cr \!\nu_3\!}. \label{propeq} 
\end{equation}

We now take the one mass scale dominance limit, and 
for convenience add
the term $\Delta / 6E \times (\rm{identity \, matrix})$ to 
the evolution matrix.
This will convert Eq. (\ref{propeq}) to,
\begin{equation}
i {d \over dx} \pmatrix{\nu_1 \cr \nu_2 \cr \nu_3}
   = {1 \over 2E} \pmatrix{0 &0 &0 \cr 0 & 0 & 0 \cr 0&0& \Delta }
   \pmatrix{\nu_1 \cr \nu_2 \cr \nu_3}. 
\end{equation}

Next, we switch to the flavor basis and add the 
effective mass term from
$e-\nu_e$ forward exchange scattering. The 
propagation equation now becomes,
\begin{equation}
i {d \over dx}\! \pmatrix{\!\nu_e\! \cr \!\nu_{\mu}\! 
\cr \!\nu_{\tau}\! }\!
   =\! {1 \over 2E} {\hat {\cal M}}^2
   \pmatrix{\!\nu_e\! \cr \!\nu_{\mu}\! \cr \!\nu_{\tau}\! }\!. 
\end{equation}
Here ${\hat {\cal M}}^2$ is the flavor-basis 
effective mass matrix in matter,
in which the extra contribution to the electron 
neutrino mass due to 
interactions with the background matter is denoted by $A$:
\begin{equation}
{\hat {\cal M}}^2 = \! U\! \pmatrix{0\! &0\! &0 \cr 0\! & 0\! & 0 
   \cr 0\!&0\!& \Delta }\!
   U^{\dag}\! + \!\pmatrix{A\!&0\!&0 \cr 0\!&0\!&0 \cr 0\!&0\!&0}\!.
\label{flavm}
\end{equation}
Here we take
$A=2{\sqrt 2} G_F N_e E$, where $N_e$ is the 
{\em net} number density of 
electrons. Note that in ${\hat {\cal M}}^2$ we 
have not included either the
diagonal (in the flavor basis) or off-diagonal 
contributions to the neutrino
effective mass matrix from neutrino-neutrino 
neutral current forward exchange
scattering (cf. \cite{fuller87} and \cite{qian95}). 
We will return to the
possible effects of these neglected terms at the 
end of this section.

In Eq. (\ref{flavm}), we take $U$ to be the 
Cabbibo-Kobayashi-Maskawa 
(CKM) matrix in the Review of Particle Properties \cite{revpart}:
\begin{equation}
U\!\!\! =\!\!\! \pmatrix{c_{12} c_{13} & s_{12} c_{13} & 
   \!\!\!s_{13} e^{-i \delta_{13}}\!\cr
   \!\!\!-\!\!s_{12} c_{23}\!\! - 
   \!\!c_{12} s_{23} s_{13} e^{i \delta_{13}}\!\!\!&
   \!\!\!c_{12} c_{23}\!\! - \!\!s_{12} s_{23} s_{13} e^{i \delta_{13}}\!\!\! &
   s_{23} c_{13} \cr \!\!\!s_{12} s_{23}\!\! - 
   \!\!c_{12} c_{23} s_{13} e^{i \delta_{13}}\!\!\! &
   \!\!-\!\!c_{12} s_{23}\!\! - \!\!
   s_{12} c_{23} s_{13} e^{i \delta_{13}}\!\!\! & c_{23} c_{13} }\!\!, \label{u} 
\end{equation}
where $c_{12} \equiv \cos \theta_{12}$, $s_{12} \equiv
\sin \theta_{12}$, and so on; and $\delta_{13}$ is 
the CP-violating phase. 
We indicate the elements of $U$ by
$U_{\alpha i}$, where $\alpha$ is a flavor index and $i$ is a mass 
eigenvalue index. In this notation, the amplitude for a neutrino to
be found in flavor eigenstate $\alpha$ is, in 
terms of the amplitudes
for the neutrino to be in the mass eigenstates $i$, 
\begin{equation}
\nu_{\alpha} = \sum_i U_{\alpha i} \nu_i.
\end{equation}
The matrix $U$ is a product of three unitary matrices, 
\begin{mathletters}
\begin{eqnarray}
U = &&U_{\theta_{23}} \,U_{\theta_{13}} \,U_{\theta_{12}},  \\ 
\nonumber \\
U_{\theta_{23}} = &&\pmatrix{1&0&0 \cr 0& c_{23} & s_{23} \cr
	0& -s_{23} & c_{23} }, \\
\nonumber \\
U_{\theta_{13}} = &&\pmatrix{c_{13} & 0 & s_{13} e^{-i \delta_{13}} \cr
	0 & 1 & 0 \cr
	-s_{13} e^{i \delta_{13}} & 0 & c_{13}}, \\
\nonumber \\
U_{\theta_{12}} = && \pmatrix{c_{12} & s_{12} & 0 \cr
	-s_{12} & c_{12} & 0 \cr 0&0&1}.
\end{eqnarray}
\end{mathletters}

We now proceed to simplify ${\hat {\cal M}}^2$. 
First, note that $\theta_{12}$
drops out:
\begin{equation}
  U_{\theta_{12}}  \pmatrix{0 &0 &0 \cr 0 & 0 & 0 \cr 0&0& \Delta }
  U^{\dag}_{\theta_{12}} = 
\pmatrix{0 &0 &0 \cr 0 & 0 & 0 \cr 0&0& \Delta }.
\end{equation}
Further, we can rotate away the angle $\theta_{23}$ by means 
of a unitary transformation: 
\begin{equation}
 {\hat{ \cal M}}_{\rm{rot}}^2 = 
U^{\dag} _{\theta_{23}} {\hat {\cal M}}^2 U_{\theta_{23}}.
\end{equation}
This rotates the flavor basis, but  
only the $\nu_{\mu}$ and $\nu_{\tau}$ amplitudes are mixed.
The $\nu_e$ amplitude, which we are most 
interested in, remains unchanged.
The background matter contribution to the effective mass matrix 
(the matrix containing $A$) is seen
from Eq. (\ref{flavm}) to be unchanged as well.

After subtracting $\Delta / 2 \times$(identity matrix) from
${\hat {\cal M}}_{\rm{rot}}^2$, we are
left with our final mass matrix in matter, 
${\hat {\cal M}}_{\rm{fin}}^2$:
\begin{equation}
{\hat {\cal M}}_{\rm{fin}}^2 = {1 \over 2} 
\pmatrix{2A - \Delta \cos 2\theta_{13} &
   0 & \Delta \sin 2\theta_{13}\, e^{-i \delta_{13}} \cr
   0 & -\Delta & 0 \cr
   \Delta \sin 2\theta_{13} \,
e^{i \delta_{13}} & 0 & \Delta \cos 2\theta_{13}}.
\end{equation}
Note that the first and third rows of this 
matrix effectively are decoupled 
into a $2 \times 2$ matrix, which can be 
recognized as that which arises in the two-neutrino mixing case
(apart from the phase factors $e^{\pm i \delta_{13}}$). The 
eigenvalues of this decoupled part of ${\hat {\cal M}}_{\rm{fin}}^2$ 
are identical to the eigenvalues in the two-neutrino mixing case. 

At the
neutrinosphere, the $| \nu_e \rangle$ eigenstate coincides with the heaviest
mass eigenstate in matter, $| \nu_3^m \rangle$, 
due to the high matter density.
The $\nu_e$ survival probability is thus 
given in the adiabatic limit by, 
\begin{equation}
   P(\nu_e \rightarrow \nu_e) = 
|\langle \nu_e | \nu_3^m \rangle_{\rm{WFO}}|^2.
\end{equation}
A similar expression obtains in the two-mixing 
case, with $| \nu_2^m \rangle$ 
being the heavier mass eigenstate in matter. 
Direct comparison of the
eigenvectors obtained from the decoupled $2 \times 2$ 
part of  ${\hat {\cal M}}_{\rm{fin}}^2$ with the eigenvectors 
obtained in the 
two-neutrino mixing case reveals that,
\begin{equation}
  \langle \nu_e | \nu_3^m \rangle_{3-\rm{flavor, \, OMSD}} = 
  e^{-i \delta_{13}} \langle \nu_e | \nu_2^m \rangle_{2-\rm{flavor}}.
\end{equation}
Of course, the factor $e^{-i \delta_{13}}$ will 
disappear in the survival probability. 
Thus, with our assumptions and approximations we have,
\begin{equation}
 P(\nu_e \rightarrow \nu_e)_{3-\rm{flavor,\,OMSD}}=
 P(\nu_e \rightarrow \nu_e)_{2-\rm{flavor}}.
\end{equation}

We now comment on nonadiabaticity. 
The assumption of an ``adiabatic''
resonant transformation means that a 
neutrino in a given mass eigenstate
remains in that mass eigenstate throughout 
its propagation. Near a
resonance, though, there is a probability 
that the neutrino will cross
onto another mass eigenstate track 
(see Fig. 1 of Ref. \cite{bethe}). 
Since in the OMSD limit of the three-neutrino mixing case
there is only one resonance between two mass 
eigenstates, the effects of
nonadiabaticity will be the same as in the 
two-neutrino mixing case.

Next we consider the effects of terms in the neutrino
propagation matrix due to neutrino-neutrino forward scattering. 
The explicit calculations in Refs. \cite{qian95,sigl95} of the
two-neutrino mixing case showed that one of the effects of 
these terms is to slightly
extend the excluded region from $\delta m^2 
\approx 4$ eV$^2$ down
to $\delta m^2 \approx 1.5$ eV$^2$ 
(see Fig. 9 of Ref. \cite{qian95}).
This is due to the nonlinear nature of the 
problem: as flavor conversion
takes place, the neutrino background is altered. 
Specifically, for small $\delta m^2$, the 
neutrino background evolves
in such a way as to pull the resonance positions of neutrinos
closer to the neutrinosphere. This behavior 
can bring the resonance positions 
associated with smaller $\delta m^2$ inside 
the weak freeze-out radius.
In turn, resonances inside the weak 
freeze-out radius can lead to a decrease
in the $n/p$ ratio. The net result in 
Refs. \cite{qian95,sigl95} is that the lower 
limit on $\delta m^2$ in the parameter 
space region excluded by $r$-process
considerations is {\em decreased}.

While we have not done explicit calculations, 
we argue that the effect of the
neutrino background will be similar in the present case. 
In the OMSD limit we consider here, the $\nu_e$ can mix 
with two other flavors of neutrinos, but only 
{\em one} of the mass differences is 
large enough to cause resonant flavor 
conversion inside the weak freeze-out 
radius of the supernova. Therefore, in the natural neutrino 
mass hierarchy considered here, 
we expect that allowing a third neutrino will 
not cause significant further
evolution of the background. This is because the additional 
neutrino mass eigenvalue difference
is too small to bring about extra flavor conversion of the $\nu_e$ 
\cite{note}.

In the natural scheme considered here, 
we can apply limits obtained in the two-neutrino 
mixing case, and conclude
that the necessity of a neutron-rich hot bubble 
environment for $r$-process
nucleosynthesis places limits on $\sin^2 2\theta_{13}$ for $\Delta
\gtrsim 1.5$eV$^2$. Of course, this conclusion 
is valid only if the post core-bounce 
supernova environment is in fact the site of 
origin of the $r$-process elements.

\section{An attempt at a three-neutrino mixing solution}
\label{sec:solution}

We seek a set of neutrino masses and mixing 
angles consistent with the
following:
  1. Accelerator and reactor data, including LSND;
  2. Atmospheric neutrinos;
  3. Solar neutrinos;
  4. Supernova $r$-process nucleosynthesis;
  5. Cold+hot dark matter models.
We attempt to do this by invoking a vacuum 
neutrino oscillation interpretation 
for both LSND and atmospheric neutrinos
with $\Delta \equiv \Delta_{13} \approx \Delta_{23} 
\approx 0.3 $ eV$^2$,
while we solve the solar neutrino problem by 
employing the MSW effect 
with $\Delta_{12} \approx 10^{-5}$ eV$^2$.

First we consider the accelerator and reactor data. The authors of 
Ref. \cite{fogli95} have made a thorough 
reanalysis of the exisiting accelerator and 
reactor data, the results of
which we employ here. For two-flavor
vacuum neutrino oscillations, the survival 
probability $P$ is given by,
\begin{equation}
P=1-\sin^2 2\theta \, \sin^2 \left(1.27 { {L \Delta }\over E} \right),
\end{equation}
where $L$ (in km) is the path length for a 
neutrino initially in a flavor
eigenstate at $L=0$, $E$ the neutrino energy in
MeV, and $\Delta $ is in eV$^2$. The two-flavor 
mixing angle is $\theta$.
In the OMSD limit, a reinterpretation of vacuum 
neutrino oscillations
in terms of three-neutrino mixing is possible. 
In this limit, we can make the
following correspondence between the two-flavor 
mixing angle and the 
three-neutrino mixing CKM matrix elements:
\begin{mathletters}
\begin{eqnarray}
\sin^2 2\theta \Leftrightarrow &4 |U^2_{\alpha 3}| |U^2_{\beta 3}| \; \;
  &(\rm{appearance});   \label{corres}\\
\sin^2 2\theta \Leftrightarrow &4 |U^2_{\alpha 3}| 
(1-|U^2_{\alpha 3}|) \; \;
  &(\rm{disappearance});
\end{eqnarray}
\end{mathletters}
for appearance and disappearance experiments, respectively. 

Eq. (\ref{corres}) shows how the solar 
neutrino problem and the LSND experiment can be
simultaneously explained. To solve the solar neutrino 
deficit, our neutrino mixing scheme requires 
an MSW resonant flavor conversion. In this conversion,
 the mass eigenstate 
corresponding most closely to $\nu_e$ deep in the interior of
the sun corresponds more closely to $\nu_{\mu}$ upon leaving
the sun. However, a two-flavor neutrino oscillation 
interpretation of the LSND signal would require 
$\nu_e$-$\nu_{\mu}$ oscillation parameters inconsistent
with those required for this MSW solution to the solar neutrino
problem. A three-neutrino mixing interpretation 
provides the resolution: Eq. (\ref{corres}) shows that the appearance 
of ${\overline \nu}_e$ in a ${\overline \nu}_{\mu}$ beam, as in 
the LSND experiment, is mediated by transitions involving
the third mass eigenstate, which 
most closely corresponds to the
$\nu_{\tau}$. The utility of these so-called ``indirect neutrino
oscillations'' for simultaneously explaining the solar neutrino
problem and certain accelerator experiments was recently noted in Ref.
\cite{babu95}.

From Eq. (\ref{u}) we have
\begin{mathletters}
\begin{eqnarray}
|U^2_{e3}| = &&\sin^2 \theta_{13}, \\
   |U^2_{\mu 3}| = &&\cos^2 \theta_{13} 
\sin^2 \theta_{23}, \\
   |U^2_{\tau 3}| = &&\cos^2 \theta_{13} \cos^2 \theta_{23}. 
\end{eqnarray}
\end{mathletters}
In Ref. \cite{fogli95}
the parameter space is displayed in a ($\log \,\tan^2 \theta_{23}$, 
$\log \, \tan^2 \theta_{13}$) plane, 
with a different plot necessary for each value of
$\Delta $. We shall employ this method of 
displaying neutrino mixing parameter space as well.

For atmospheric neutrinos, we here use only 
sub-Gev data (c.f. our comments
on multi-GeV data in Sec. \ref{sec:intro}). Results of 
atmospheric neutrino experiments
are often reported as the ``ratio of ratios'' $R$ ,
\begin{equation}
R = { {(\nu_{\mu} / \nu_e)_{\rm{data}} } \over 
  {(\nu_{\mu} / \nu_e)_{\rm{Monte \; Carlo}} } }. 
\end{equation}
In the sub-GeV range, for three-neutrino mixing $R$ 
is given by (\cite{acker94} and references therein),
\begin{equation}
R = { {P_{\mu \mu} + r P_{\mu e}} \over {P_{ee} + 
r^{-1} P_{\mu e} } }.
\end{equation}
Here $r$ is a particular ratio of electron- to 
muon-type neutrinos, which 
we take to be $r = 0.49$ \cite{atrev}. The mass 
scale we are interested in implies an averaging of 
the oscillation factors:
\begin{equation}
\sin^2 \left(1.27 { {L \Delta}\over E} \right) \approx {1\over 2}, 
\end{equation}
so that $R$ becomes independent of $\Delta$ for 
large enough $\Delta$.

Part of our putative ``solution to everything'' appears 
in Fig. 1. In this figure
we show constraints on four ($\log \,\tan^2 \theta_{23}$, 
$\log \, \tan^2 \theta_{13}$) panels. Each panel 
corresponds to a different
value of $\Delta$ ($\Delta = 0.2$ eV$^2$, 0.3 eV$^2$, 
0.4 eV$^2$, 0.5 eV$^2$).
The accelerator and
reactor data in Fig. 1 are given at 95\% C.L. Except for 
LSND, the data 

\epsfysize=8cm \epsfbox{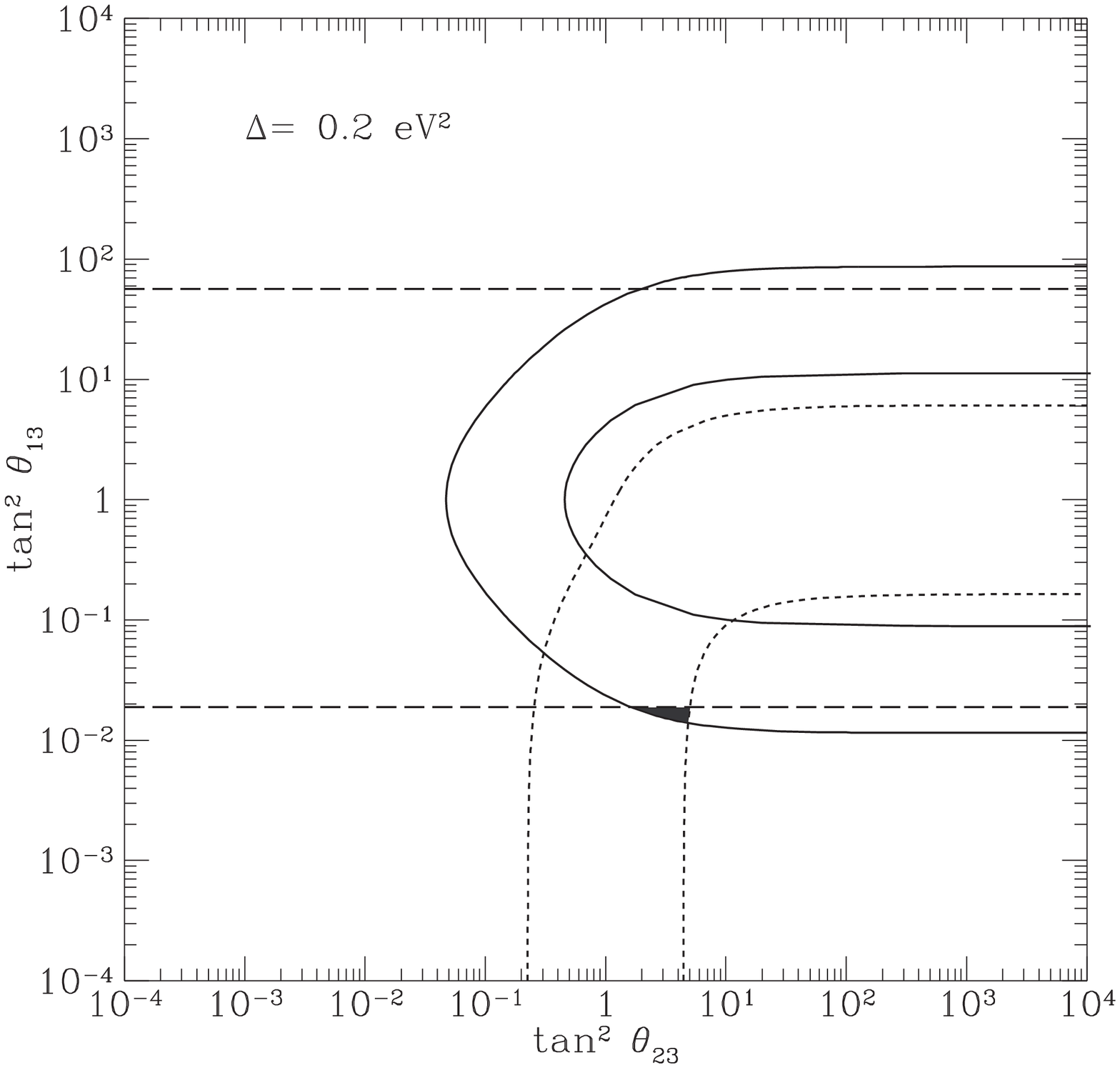}

\epsfysize=8cm \epsfbox{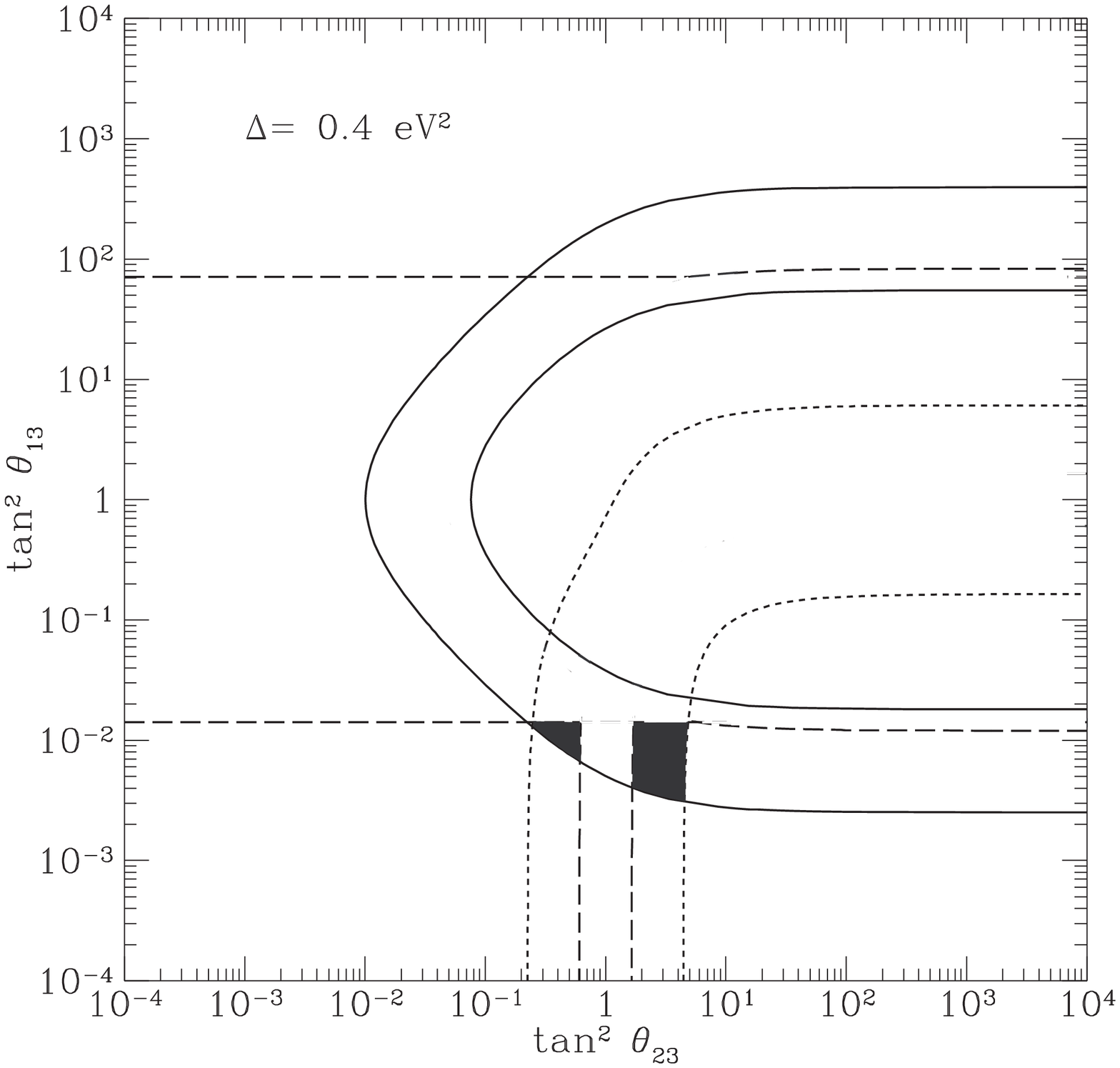}

\begin{figure}
\widetext
\caption{Allowed regions of the mixing angles $\theta_{13}$ and $\theta_{23}$,
	for four values of the dominant mass difference. 
	The region between the
	solid lines is the 95\% C.L. detection of LSND. The region inside
	the long dashed lines is excluded by accelerator/reactor 95\%
	C.L. limits. The region enclosed by the short dashed lines 
	represents a solution to the sub-GeV atmospheric neutrino data, with
	$R = 0.7$. See text for discussion.}
\narrowtext
%\label{}
\end{figure}

\noindent are taken from
the reanalysis of accelerator and reactor experiments 
in Ref. \cite{fogli95}.
The LSND 95\% C.L. data are taken from an early 
preprint of the Ref. 
\cite{lsnd} group.  To delineate the band 
allowed by atmospheric neutrinos in Fig. 1 we 
have taken $R=0.7$; the band shrinks 
if smaller values of $R$ are employed. Note that 
$R=0.7$ is not the central reported value. However, 
we have employed this high $R$ value  
in the interest of discovering what the data 
might accommodate .
This value of $R$ is in the vicinity of the upper 
limits on this quantity 
allowed by the Kamiokande 
\cite{kamiokande} and IMB \cite{imb} sub-GeV data.

The $\Delta=0.2$ eV$^2$ panel of Fig. 1 shows 
where the lower 

\epsfysize=8cm \epsfbox{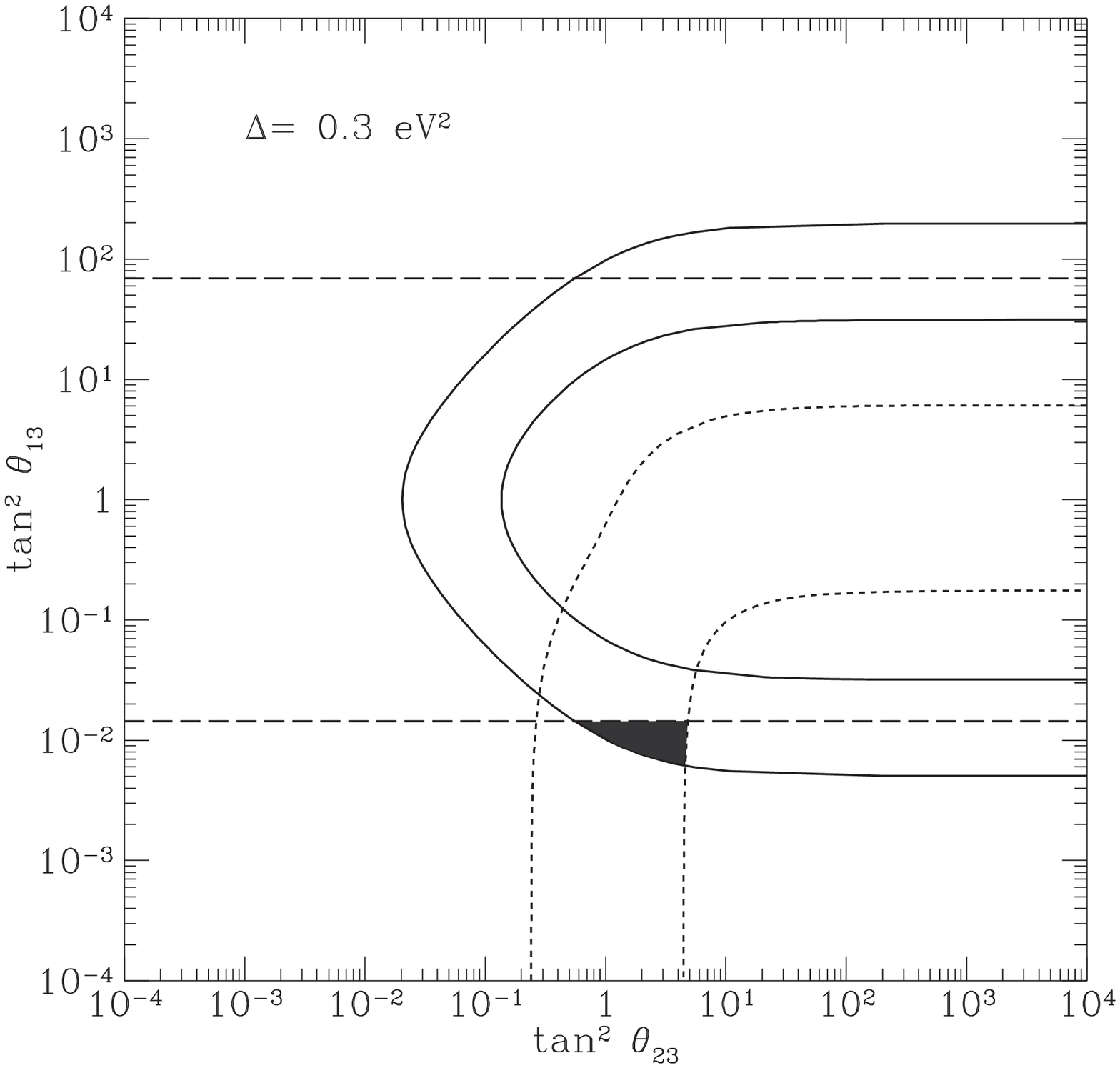}

\epsfysize=8cm \epsfbox{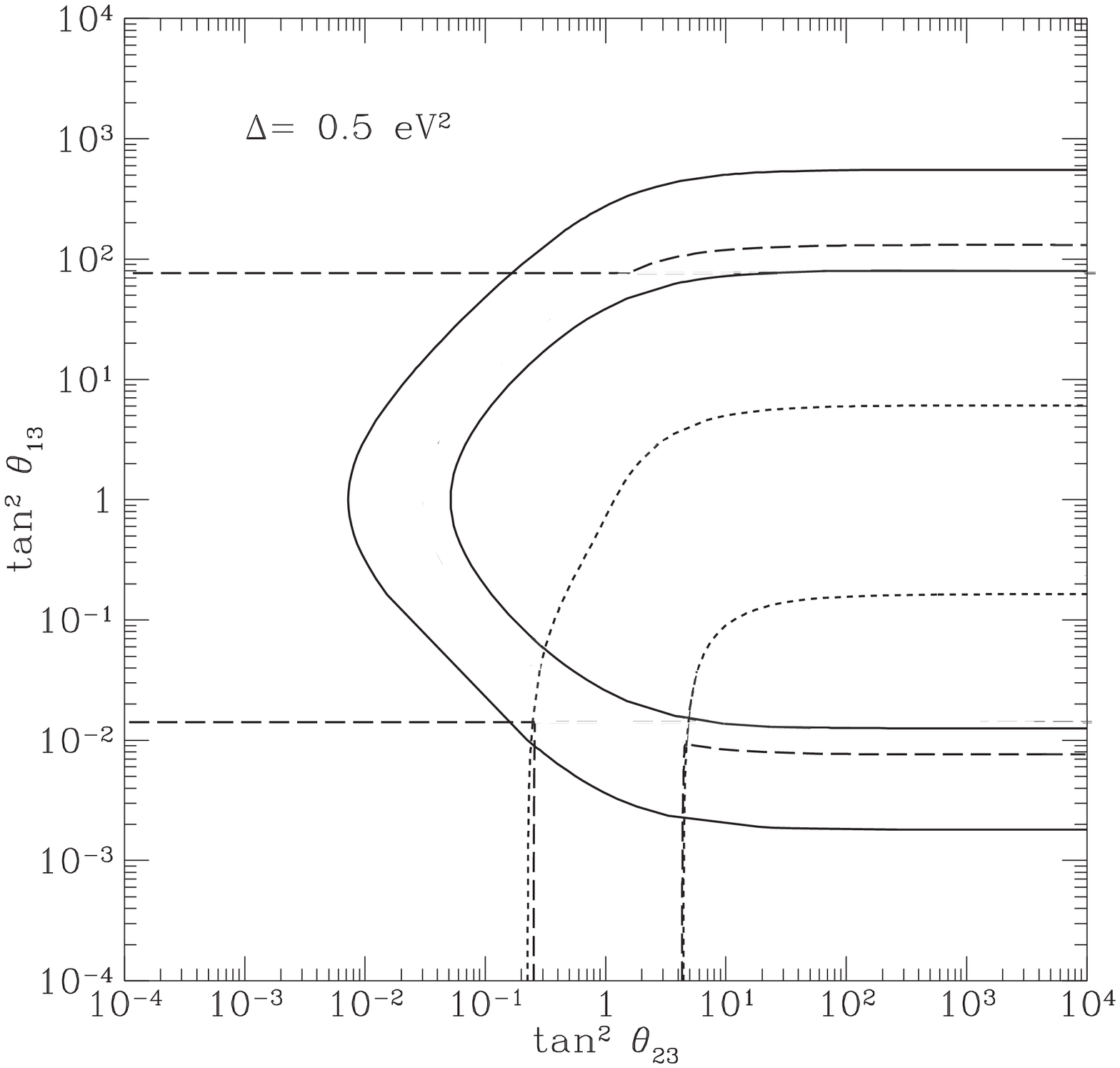}
\vskip 2.2cm

\noindent bound 
on $\Delta$ originates: below $\Delta
\approx 0.2$ eV$^2$, LSND begins to become 
incompatible with $\nu_e$
disappearance constraints. These $\nu_e$ 
disappearance limits come from 
reactor experiments, particularly the
Bugey reactor experiment \cite{bugey}. The 
$\Delta=0.5$ eV$^2$ panel of
Fig. 1 shows 
where the upper bound on $\Delta$ comes from: above $\Delta
\approx 0.4$ eV$^2$,\  $\nu_{\mu}$ disappearance experiments 
(in particular CDHSW \cite{cdhsw}) rule out most 
of the atmospheric neutrino solution.
The panels of Fig. 1 with $\Delta=0.3$ eV$^2$ and 
$\Delta=0.4$ eV$^2$
 show the allowed regions of parameter space. We note that
$\Delta \approx 0.3$ eV$^2$ is perhaps the safest solution. 
With the choice $\Delta \approx 0.4$ eV$^2$,
the central part of the atmospheric neutrino band 
partially overlaps the 
disallowed parameter space region of CDHSW.  
This is significant, because
the central part of the atmospheric neutrino 
band is typically the part that would
remain if a lower value of $R$ were to be chosen.
For solar neutrinos, we invoke an MSW solution 
using the parameters
$\Delta_{12}$ and $\theta_{12}$. These quantities 
dropped out in the
above analysis of accelerator/reactor experiments, 
atmospheric neutrinos,
and the supernova $r$-process, due to our use of 
the OMSD limit. However, 
$\Delta_{12}$ and $\theta_{12}$ are important near the
mass level crossing that occurs in the sun. 
This is because of the extra effective mass acquired 
by the $\nu_e$.  
Using the 
decoupled two-neutrino solution obtained in Ref. 
\cite{kuo87} (valid
for the well-separated mass scales which we 
have here), the authors of Ref.
\cite{fogli94} obtain solutions to the solar neutrino problem. Their
Figs. 4 and 6 show the allowed areas in the ($\Delta_{12}$, $\sin^2
2\theta_{12} / \cos 2\theta_{12}$) plane for $\sin^2 \theta_{13} =
0.0$ and $\sin^2 \theta_{13}= 0.1$, respectively. 
It is apparent that the ``small angle'' 
solution is essentially unchanged for these 
different values of $\sin^2
\theta_{13}$. From our Fig. 1, we see that our 
putative solution has $\sin^2
\theta_{13} \approx 10^{-2}$, so we must adopt 
the small angle solution 
seen in Figs. 4 and 6 of Ref. \cite{fogli94}: $\Delta_{12} \approx
7 \times 10^{-6}$ eV$^2$; $\sin^2 \theta_{12} 
\approx 2 \times 10^{-3}.$
The CP-violating phase factor only appears in 
terms that vanish in the 
approximations used in Refs. \cite{kuo87} and \cite{fogli94}.

For massive neutrinos to be of use for cold+hot 
dark matter models,
it is desirable to have the neutrino masses add up to about 5 eV 
\cite{primack95}.
Given our value of $\Delta$, this implies $m_1 \approx m_2 \approx
1.5$ eV, and $m_3 \approx 2$ eV. This may appear to put the 
$\nu_e$ Majorana
mass in conflict with the limit $m_{\nu_e} \lesssim 0.7$ eV from 
neutrinoless double beta decay \cite{doublebeta}. 
However, this is not necessarily a serious
problem. As discussed in Ref. \cite{raffelt95}, a more realistic limit 
on the $\nu_e$ Majorana mass may
be around 1.4 eV. Also, perhaps cold+hot dark 
matter models could
work with the sum of the neutrino masses being a 
little less than 5 eV.
Finally, if the neutrinos have Dirac masses,
the limit from neutrinoless double beta decay does 
not apply at all. 

We have found a fairly unique set of neutrino masses and
mixing angles satisfying the constraints listed at 
the beginning of this section. We note that the 
value of $\Delta$ we have
been led to is safe from the perspective of the 
$r$-process. Additionally,
in the absence of a significant net lepton number in the universe, 
neutrino oscillations with these parameters should
have essentially no effect on the outcome of 
big bang nucleosynthesis
\cite{shi93}.

However, this natural solution does not look very 
promising. The panels in Fig. 1
contain 95\% C.L. data (with the exception of the 
atmospheric neutrino
``ratio of ratios'' $R$, for which confidence levels 
are not readily assignable).
At 90\% C.L., the LSND detection band shrinks, 
while the excluded regions from the
other reactor/accelerator limits expand. This  solution then exists 
essentially as a point in the ($\log \,\tan^2 \theta_{23}$, 
$\log \, \tan^2 \theta_{13}$) plane, and this only 
at $\Delta \approx 0.3$ eV$^2$.
The existence of a three-neutrino mixing solution, 
in the OMSD limit and
satisfying the five points at the beginning of this 
section, is therefore fragile at best.

It could also be argued that this ``natural'' solution 
is not so natural after all.
For example, we have taken $\Delta_{13} \approx 
\Delta_{23} \gg \Delta_{12}$,
in analogy with the hierarchy of mass differences 
of the charged leptons.
For the charged leptons, this hierarchy of 
mass {\em differences} arises naturally
from the hierarchy of the {\em masses themselves}, 
i.e. $m^2_e \ll m^2_{\mu} 
\ll m^2_{\tau}$.
However, the offset of neutrino masses from zero 
required to make the neutrino
masses sum to about 5 eV for cold+hot dark matter 
models causes the absolute 
masses to have roughly similar magnitudes, in 
contrast to the masses of the 
charged leptons.

Another unnatural feature of this putative ``natural'' 
solution is that 
several of the off-diagonal 
elements of the mixing matrix $U$ have relatively 
large magnitudes. 
This is in contrast to the quark mixing case \cite{revpart}. It has
been recognized previously that a three-neutrino 
oscillation explanation of
the LSND experiment requires this unusual feature 
\cite{bilenky95}. 
Large off-diagonal
terms will generally be present whenever oscillation 
probabilities are
large, and neutrino oscillation explanations of 
the atmospheric neutrino 
anomaly and the LSND data both invoke 
relatively large oscillation
probabilities. 

Note from Fig. 1 that the LSND oscillation signal significantly
restricts the ``natural'' solution. If the LSND data are interpreted
as yielding only an upper limit as in Ref. \cite{hill95}, the 
atmospheric neutrino anomaly can be solved with any value
of the mixing angle $\theta_{13}$ such that $\tan^2 \theta_{13}
\lesssim 10^{-2}$. The lower limit of $\Delta \approx 0.2$ eV$^2$
would also disappear in this case, allowing the experimentally more 
comfortable value of $\Delta \approx 10^{-2}$ eV$^2$ to be
used to solve the atmospheric neutrino problem. Even without
the LSND detection, however, the simultaneous solution of all
the remaining constraints would still involve the unnatural 
features mentioned above: nearly degenerate neutrino masses
and large off-diagonal elements in the neutrino mixing matrix,
in contrast with the properties of the charged leptons.

\section{Conclusion}
\label{sec:concl}

Some models which have sought to account 
for currently  available clues
about neutrino properties
have employed neutrino oscillations with sterile neutrinos, 
and/or an inverted mass hierarchy 
\cite{neumodels,raffelt95,chun95}. 
These devices are required when 
two-flavor interpretations of various physical effects are ``stitched
together'' to make a consistent composite model. 
Here we have sought a ``natural'' three-neutrino mixing
scheme---without sterile neutrinos or an 
inverted mass hierarchy---that
satisfies:
  1. Accelerator and reactor data, including LSND;
  2. Atmospheric neutrinos;
  3. Solar neutrinos;
  4. Supernova $r$-process nucleosynthesis;
  5. Cold+hot dark matter models.
Along the way we have argued that, under 
some circumstances, putative supernova 
$r$-process nucleosynthesis bounds on
two-neutrino flavor mixing can be applied 
directly to three-neutrino mixing
in the case of one mass scale dominance.

We have found a possible ``natural'' solution, 
and it is quite restricted. 
The mass differences in this putative solution
are $\Delta_{12} \approx 7 \times 10^{-6}$ 
eV$^2$ (for an MSW solution to the
solar neutrino problem), and $\Delta_{13} \approx 
\Delta_{23} \approx 0.3$ eV$^2$ (for LSND and 
atmospheric neutrinos). 
Here $\Delta_{ji} \equiv |m^2_j - m^2_i|$.
To be of use to cold+hot dark matter models, we take the 
masses themselves
to be $m_1 \approx m_2 \approx 1.5$ eV, and $m_3 
\approx 2$ eV. The mixing
angles we require are $\sin^2 \theta_{12} \approx 2 
\times 10^{-3}$,
$\sin^2 \theta_{13} \approx 10^{-2}$, and $\sin^2 
\theta_{23} \sim 0.5$
(see Figs. 1a-1d for $\theta_{13}$ and 
$\theta_{23}$, and Figs. 4 and 6 of Ref.
\cite{fogli94} for $\theta_{12}$). While this 
``natural'' solution exists, 
it is rather fragile and
has some arguably unnatural features, as discussed in
Sec. \ref{sec:solution}.

Perhaps the main difficulty with this scheme is finding a common 
mass difference suitable for a vacuum neutrino 
oscillation solution for
both LSND and the zenith-angle dependance
of the Kamiokande multi-GeV atmospheric 
neutrino data. As the statistics for both of these
experiments are not compelling at this stage, 
future results may shed
light on the matter. Super-Kamiokande \cite{superkam} 
will have much better statistics for the atmospheric neutrino
deficit. Furthermore, the LSND experiment continues and may
provide better statistics in the future.
The KARMEN experiment \cite{armbruster95}, 
which probes regions of parameter space
similar to LSND, hopefully will also report further results.

Proposed terrestrial experiments could 
definitively eliminate the solution
presented in the last section. New reactor 
experiments at San Onofre and
Chooz will have increased sensitivity to 
$\delta m^2$, but apparently will
not have increased sensitivity to the oscillation 
probability at the mass
difference we are interested in \cite{sanonofre}. 
Conversely, the CHORUS and
NOMAD experiments have high sensitivity to the 
oscillation probability, but will
probably not reach small enough $\delta m^2$ 
to convincingly eliminate
the putative ``natural'' solution \cite{cernexp}. 
However, proposed long-baseline accelerator 
experiments will probe the entire region of 
parameter space favored by 
the atmospheric neutrino sub-GeV data 
\cite{fogli95,michael95} and could thus
provide the crucial test.

Assuming the validity of the data and astrophysical arguments 
we have attempted to satisfy, the convincing experimental 
elimination of the last remaining ``natural'' solution 
presented here would be an important development. It would be
significant evidence for the existence of sterile neutrinos and/or
a neutrino mass hierarchy of a different nature than that
of the charged leptons (i.e. an ``inverted'' hierarchy, in which the 
mass eigenvalue most closely associated with $\nu_e$ is heavier
than those associated with $\nu_{\mu}$ or 
$\nu_{\tau}$; or a hierarchy
in which no one neutrino mass eigenvalue difference 
dominates the other
neutrino mass eigenvalue differences).

\section*{Acknowledgements}  
We thank the organizers of the 1995 Santa Fe Workshop
on massive neutrinos for providing a stimulating environment
where this work was begun. We also thank A. B. Balantekin, 
D. Caldwell, Y.-Z. Qian, and W. Vernon for
useful discussions.
This work was supported by the National 
Science Foundation through NSF
Grant PHY-9503384, and by a NASA Theory Grant.

% now the references. delete or change fake 
%bibitem. delete next three
%   lines and directly read in your .bbl file if you use bibtex.

% Fill in the caption in the braces of the \caption{} 
% command. Put the label

% \begin{table}
% \caption{}
% \label{}
% \end{tabular}
% \end{table}


\begin{references}
\bibitem{note}Note, however, that this conclusion 
	might be altered if the $\nu_{\mu}$
	and $\nu_{\tau}$ were nearly degenerate, i.e. if 
	$\Delta_{12} \approx \Delta_{13} \gg
	\Delta_{23}$.
\end{references}
\end{document}